\documentclass[pra,twocolumn,amsmath,amssymb,superscriptaddress,showpacs,floatfix]{revtex4-1}
\usepackage{graphicx}
\usepackage{amsmath}
\usepackage{soul,xcolor}
\usepackage{color}
\usepackage{placeins}
\usepackage{braket}
\usepackage[normalem]{ulem}

\usepackage{mathtools}

\definecolor{dgreen}{rgb}{0.0, 0.5, 0.0}

\newcommand{\veck}{\mathbf k}

\newcommand{\vecp}{\mathbf p}
\newcommand{\vecq}{\mathbf q}

\newcommand{\vecr}{\mathbf r}

\newcommand{\vecR}{\mathbf R}
\renewcommand{\braket}[2]{\langle #1 | #2 \rangle}

\newcommand{\ded}{\hat d^\dagger}
\newcommand{\de}{\hat d}
\newcommand{\bed}{\hat b^\dagger}
\newcommand{\be}{\hat b}
\newcommand{\Bed}{\hat B^\dagger}
\newcommand{\Be}{\hat B}
\newcommand{\ad}{\hat a^\dagger}
\renewcommand{\a}{\hat a}

\begin{document}

\preprint{APS/123-QED}

\title{Theory of excitation of Rydberg polarons in an atomic quantum gas}
\author{R. Schmidt }
\email{corresponding author: richard.schmidt@cfa.harvard.edu}
\affiliation{
ITAMP, Harvard-Smithsonian Center for Astrophysics, Cambridge, MA 02138, USA
}%
\affiliation{
Department of Physics, Harvard University, Cambridge, MA 02138, USA
}%
\affiliation{
 Institute of Quantum Electronics, ETH Z\"urich, CH-8093 Z\"urich, Switzerland
}

\author{J.\,D. Whalen}
\author{R. Ding}
\affiliation{
 Department of Physics \& Astronomy, Rice University, Houston, TX 77251, USA
}
\author{F. Camargo}
\affiliation{
 Department of Physics \& Astronomy, Rice University, Houston, TX 77251, USA
}

\author{G. Woehl Jr.}
\affiliation{
 Department of Physics \& Astronomy, Rice University, Houston, TX 77251, USA
}
\affiliation{
 Instituto de Estudos Avan\c{c}ados, 12.228-001 S\~{a}o Jos\'{e} dos Campos, S\~{a}o Paulo, Brazil
}

\author{S. Yoshida} 
\author{J. Burgd\"orfer}
\affiliation{%
Institute for Theoretical Physics, Vienna University of Technology, Vienna, Austria, EU
}%

\author{F.\,B. Dunning}
\affiliation{
 Department of Physics \& Astronomy, Rice University, Houston, TX 77251, USA
}

\author{E. Demler }
\affiliation{
Department of Physics, Harvard University, Cambridge, MA 02138, USA
}%

\author{H.\,R. Sadeghpour }
\affiliation{
ITAMP, Harvard-Smithsonian Center for Astrophysics, Cambridge, MA 02138, USA
}%

\author{T.\,C. Killian}%
\affiliation{
 Department of Physics \& Astronomy, Rice University, Houston, TX 77251, USA
}

\date{\today}

\begin{abstract}
{We present a quantum many-body description of the excitation spectrum of Rydberg polarons in a Bose gas. The many-body Hamiltonian is solved with a functional determinant approach, and we extend this technique to describe Rydberg polarons of finite mass. Mean-field and classical descriptions of the spectrum are derived as approximations of the many-body theory. The various approaches are applied to experimental observations of polarons created by excitation of Rydberg atoms in a strontium Bose-Einstein condensate.
}
\end{abstract}

\maketitle




\section{Introduction}

When an impurity is immersed in a polarizable medium, the collective response of the medium can form quasi-particles, labelled as polarons, which describe the dressing of the impurity by excitations of the background medium. Polarons play important roles in the conduction in ionic crystals and polar semiconductors \cite{dpe82}, spin-current transport in organic semiconductors \cite{wak14}, dynamics of molecules in superfluid helium nanodroplets \cite{Schmidt2015ang, SchmidtLem2016, lemeshko2017}, and collective excitations in strongly interacting fermionic and bosonic ultracold gases \cite{swa09,hvk16,jws16}.

In this paper, which accompanies the publication heralding the observation of Rydberg Bose polarons \cite{SrPRL2017}, we present details of calculations and the interpretation of this observation as a new class of Bose polarons, formed through excitation of Sr($5sns$\,$^3S_1$) Rydberg atoms in a strontium Bose-Einstein condensate (BEC). We begin with a general outline of different polaron Hamiltonians, and construct the Rydberg polaron Hamiltonian used in this work. The spectral response function in the linear response limit is derived and the many-body mean field shift of the spectral response is  obtained. The mean field theory particularly fails to describe the limits of small and large detuning, where the detailed description of quantum few- and many-body processes is particularly relevant. These processes are fully accounted for by the bosonic functional determinant approach (FDA) \cite{ssd16} that solves an extended Fr\"{o}hlich Hamiltonian for an impurity in a Bose gas. In the frequency domain, the FDA predicts a gaussian shape for the intrinsic spectrum, which is a hallmark of Rydberg polarons.   A classical Monte Carlo simulation \cite{sln16}  which reproduces the background spectral shape is shown to  miss spectral features arising from quantization of bound states. Agreement between  experimental results and FDA theory for  both the observed few-body molecular spectra and the many-body polaronic states is excellent.

 In the companion paper \cite{SrPRL2017}, we provide experimental evidence for the observation of polarons created by excitation of Rydberg atoms in a strontium Bose-Einstein condensate, with an emphasis on the determination of the excitation spectrum in the absence of density inhomogeneity. Here we provide a detailed discussion of the theoretical methods and experimental analysis.

\section{Bose Polaron Hamiltonians}
With ultracold atomic systems, various quantum impurity models can be studied in which an impurity interacts with a bosonic bath. Here we focus on  models that follow from the general Hamiltonian  describing an impurity of mass $M$  interacting with a gas of weakly interacting bosons of mass $m$:
\begin{eqnarray}\label{appHFull}
\hat H&=&\sum_\vecp \epsilon_\vecp^I\ded_\vecp\de_\vecp
+\sum_\veck \epsilon_\veck \ad_\veck\a_\veck
+\frac{g_{bb}}{2\cal{V}}\sum_{\veck\veck'\vecq}\ad_{\veck'+\vecq}\ad_{\veck-\vecq}\a_{\veck'}\a_{\veck}\nonumber\\
&+&\underbracket{\frac{1}{\cal{V}}\sum_{ \veck\veck'\vecq}V(\vecq)\ded_{\veck'-\vecq}\de_{\veck'}\ad_{\veck+\vecq}\a_{\veck}}_{ \hat H_{IB}}.
\end{eqnarray}
Here, the first two terms describe the kinetic energy of the impurities ($\de_\vecp$) and bosons ($\a_\veck$) with dispersion relations $\epsilon_\vecp^I=\frac{\vecp^2}{2M}$ and $\epsilon_\veck=\frac{\veck^2}{2m}$, respectively (unless explicitly stated we set $\hbar=1$). The third term accounts for the interaction between the bosons. Assuming weak coupling between bosons, the microscopic coupling constant is given by the relation $g_{bb}=\pi a_{bb}/m$, with $a_{bb}$ the s-wave scattering length describing the low-energy boson-boson interactions. The last term, $\hat H_{IB}$ describes the impurity-boson interaction in momentum space, which, in the real space, reads
\begin{equation}\label{appHIB}
\hat H_{IB}=\int d^3r d^3r' \,\,\hat n_I(\vecr')\,V(\vecr'-\vecr)\,\hat n_B(\vecr).
\end{equation}
Here $\hat n_I(\vecr)=\hat\psi^\dagger_I(\vecr)\hat\psi_I(\vecr)=\frac{1}{\cal{V}}\sum_{\veck\vecq} \ded_{\veck+\vecq}\de_\veck e^{-i\vecq \vecr}$ and $\hat n_B(\vecr)=\hat\psi^\dagger_B(\vecr)\hat\psi_B(\vecr)=\frac{1}{\cal{V}}\sum_{\veck\vecq} \ad_{\veck+\vecq}\a_\veck e^{-i\vecq \vecr}$ are the impurity and boson density, respectively.

As we consider the limit of a single impurity it is convenient to switch to the first quantized description of the impurity, which is characterized by its position and momentum operator  $\hat \vecR$ and $\hat \vecp$. The density becomes $\hat n_I(\vecr)\to \delta^{(3)}(\vecr - \hat\vecR)$ and  Eq.~\eqref{appHFull} takes the form
\begin{eqnarray}\label{appHFullFirstQuant}
\hat H&=& \frac{\hat \vecp^2}{2M}
+\sum_\veck \epsilon_\veck \ad_\veck\a_\veck
+\frac{g_{bb}}{2\cal{V}}\sum_{\veck\veck'\vecq}\ad_{\veck'+\vecq}\ad_{\veck-\vecq}\a_{\veck'}\a_{\veck} \\ \nonumber
&+&\frac{1}{\cal{V}}\sum_{ \veck\vecq}V(\vecq)e^{-i\vecq \hat \vecR}\ad_{\veck+\vecq}\a_{\veck}.
\end{eqnarray}

From this Hamiltonian various polaron models can be derived, and we briefly explain their relevance and regimes of validity. At $T=0$ a large fraction of bosons is condensed in a BEC. This condensate can be regarded as a coherent state such that the zero-momentum mode $\langle \a_\veck \rangle =\sqrt{N_0}\delta_{\veck,\mathbf 0}$ takes on a macroscopic expectation value.

Within the Bogoliubov approximation the bosonic creation and annihilation operators in Eq.~\eqref{appHFullFirstQuant} are expanded in fluctuations $\hat B_\vecp$ around this expectation value $\sqrt{N_0}$ and terms of higher than quadratic order are neglected in the resulting Hamiltonian. The purely bosonic part of the Hamiltonian can then be diagonalized by the Bogoliubov rotation
\begin{eqnarray}\label{BogTrafo}
\Be_\vecp=u_\vecp \be_\vecp + v_{-\vecp}^* \bed_{-\vecp}\,\,,\,\,\Bed_\vecp=u_\vecp^* \bed_\vecp + v_{-\vecp} \be_{-\vecp},
\end{eqnarray}
where  $\omega_\veck=\sqrt{\epsilon_\vecp(\epsilon_\vecp+2 g_{bb}\rho)}$ is the Bogoliubov dispersion relation and
$u_\vecp,v_{-\vecp}=\pm\sqrt{\frac{\epsilon_\vecp+g_{bb} \rho}{2\omega_\vecp}\pm \frac{1}{2}}$. The density of the homogeneous condensate is given by $\rho$.
The transformation \eqref{BogTrafo} yields the   Bose-impurity Hamiltonian
\begin{eqnarray}\label{appExtFroehlich1}
\hat H&=& \frac{\hat \vecp^2}{2M}
+\sum_\veck \omega_\veck \bed_\veck\be_\veck  \nonumber \\
&+&\rho V(\mathbf 0)+\underbracket{\frac{1}{\sqrt{\cal{V}}}\sum_{ \vecq}g(\vecq) e^{-i\vecq \hat \vecR}\left(\bed_{\vecq}+\be_{-\vecq}\right)}_{\text{Fr\"ohlich interaction}}
\nonumber\\
&+&\underbracket{\frac{1}{\cal{V}}\sum_{ \veck\vecq}V(\vecq)e^{-i\vecq \hat \vecR}\Bed_{\veck+\vecq}\Be_{\veck}}_{\text{extended Fr\"ohlich interaction}}
\end{eqnarray}
where we introduced the `Fr\"ohlich coupling' $g(\vecq)=\sqrt{\frac{\rho \epsilon_\vecq}{\omega_\vecq}}V(\vecq)$ and in the last term we kept the untransformed expression for notational brevity. In Eq.~\eqref{appExtFroehlich1} the  term $\rho V(\mathbf 0)= \rho\int d^3r  V(\vecr)$ describes the mean-field energy shift of the polaron  in the Born approximation. Neglecting the constant mean-field shift and the last term in Eq.~\eqref{appExtFroehlich1} one arrives at the celebrated Fr\"ohlich model \cite{Frohlich1954}:
\begin{align}\label{appFroehlich}
\hat H&= \frac{\hat \vecp^2}{2M}
+\sum_\veck \omega_\veck \bed_\veck\be_\veck
+\frac{1}{\sqrt{\cal{V}}}\sum_{ \vecq}g(\vecq) e^{-i\vecq \hat \vecR}\left(\bed_{\vecq}+\be_{-\vecq}\right)
\end{align}
In initial attempts to describe impurities in weakly interacting Bose gases, this Hamiltonian was used for systems close to Feshbach resonances \cite{Cucchietti2006,tco09}. However, as shown in \cite{RathSchmidt2013}, the Fr\"ohlich Hamiltonian alone is insufficient to describe impurities that interact strongly with atomic quantum gases (for an explicit third-order perturbation theory analysis see Ref.~\cite{Christensen2015}). Indeed the major theoretical shortcoming of Eq.~\eqref{appFroehlich} is that from this model one cannot recover the Lippmann-Schwinger equation for two-body impurity-bose scattering. Hence it fails to describe the underlying two-body scattering physics between the impurity and bosons including  molecule  formation.

Similarly the Fr\"ohlich Hamiltonian cannot account for the intricate dynamics leading to the formation of Rydberg polarons. For such strongly interacting systems the inclusion of the last term in Eq.~\eqref{appExtFroehlich1} becomes  crucial. This term accounts for pairing of the impurity with the atoms in the environment  and accounts for the detailed `short-distance' physics of the problem, which is completely neglected in the Fr\"ohlich model that is tailored for the  description of long-wave length (low-energy) physics.

So far it has been experimentally verified that the inclusion of this term is relevant for the observation of Bose polarons \cite{hvk16,jws16} where the impurity-Bose interaction can be modeled by a potential that supports only a {single}, weakly bound two-body molecular state. In the present work we encounter a new type of impurity problem where the impurity is  dressed by large sets of molecular states that  have ultra-long-range character. This yields the novel physics of Rydberg polarons that is beyond the physics of Bose polarons so far observed in experiments and discussed in the literature.  To account for the relevant Rydberg molecular physics theoretically, we  analyze the full Hamiltonian Eq.~\eqref{appExtFroehlich1}.

We note that the physics of the  Rydberg oligomer states is  different from that of Efimov states in the Bose polaron problem \cite{RathSchmidt2013,Levinsen2015,Ashida2017,Sun2017}.  While Efimov states are also multi-body bound states, they arise due a quantum anomaly of the underlying quantum field theory  \cite{Moroz2010,Floerchinger2011,Schmidt2012}.
 Indeed, the Efimov effect \cite{Efimov70,Efimov71,Efimov73} gives rise to an  infinite series of three-body  states (and also states of larger atom number \cite{Stecher2009,Stecher2010}) that respect a discrete scaling symmetry. The Efimov effect arises for resonant short-range two-body interactions in three dimensions. In previous experiments studying Bose polarons close to a Feshbach resonance it was found that these Efimov states do not strongly influence the polaron physics \cite{hvk16,jws16}. In our case the multi-molecular states are not related to the Efimov effect and dimer, trimer, etc. states are rather   comprised  of nearly independently bound two-body molecular states that feature only very weak three- and higher-body correlations. In contrast, in Efimov physics, a single impurity potentially mediates  strong correlations between  two particles in the Bose gas. Due to large binding energies of Rydberg molecular states, this induced interaction is expected to be small for Rydberg polarons \cite{Fey2016}.

\subsection{Polaron formation}

\begin{figure}[t]
\includegraphics[width=0.5\textwidth]{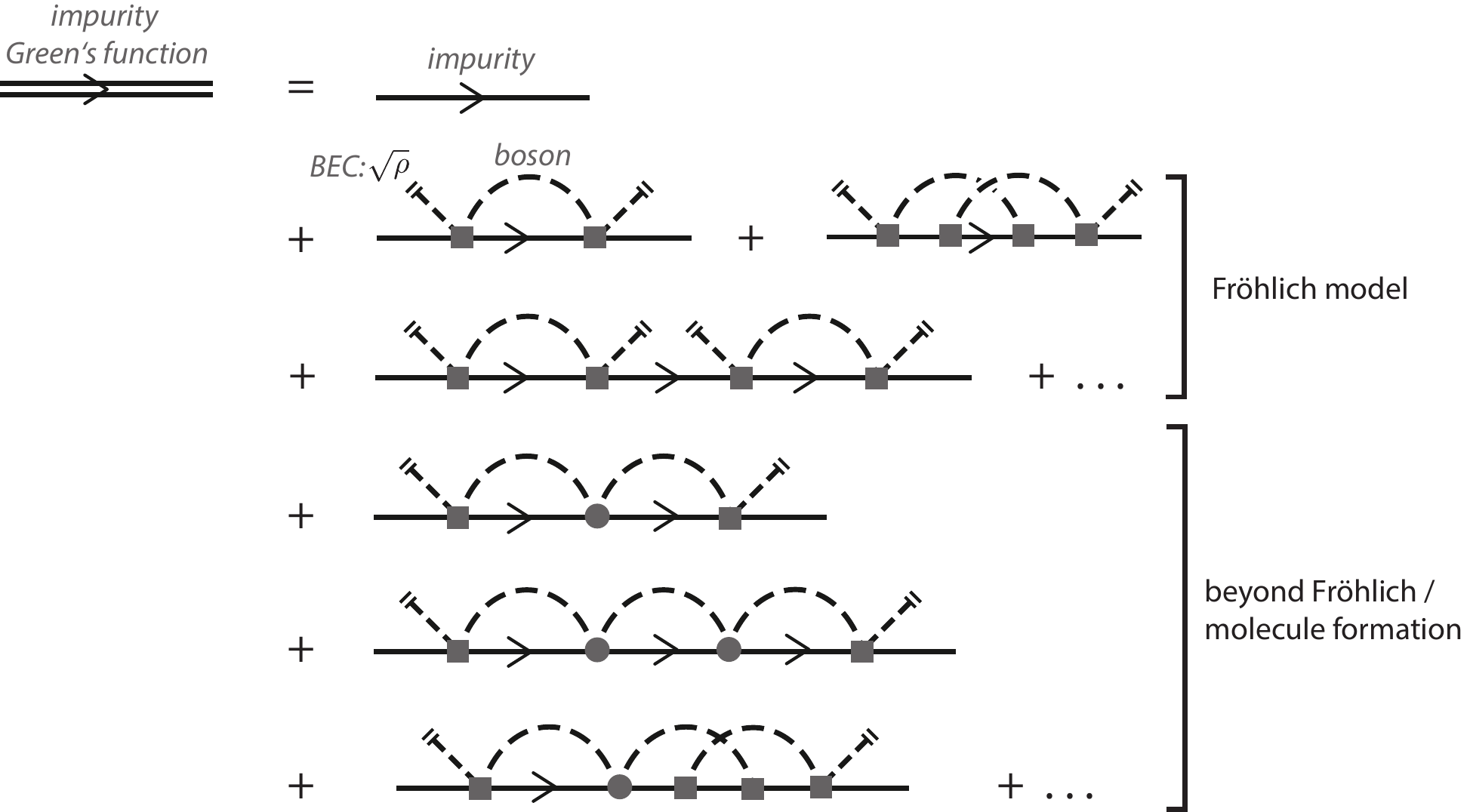}
\caption{\textbf{Illustration of polaron formation in terms of Feynman diagrams.} The impurity Green's function fully determines the impurity spectral function and hence its  absorption spectrum, mobility, and  other single particle properties. The impurity properties are changed due to the dressing by bosonic excitations signified by the self-energy corrections depicted here as diagrams. The extended Fr\"ohlich model features two types of excitations. In ``Fr\"ohlich processes'' bosons are excited   out of the BEC by scattering with the impurity (square box). Such scattering contributes a factor of $\sqrt{\rho}$. The excited bosons then reenter the BEC after an additional scattering off the impurity. A second type of process (gray disks) is ignored in the Fr\"ohlich model. Here an excited boson can scatter off the impurity repeated times and in each such scattering process it changes its momentum. These processes account for strong coupling physics in cold atomic gases including bound state formation.  }
\label{fig:Feynman}
\end{figure}

The formation of a polaron can be understood as
dressing of the impurity by excitations of the bosonic bath, which entangles the momentum of the impurity with that of the bath excitations. This can be illustrated with a simple wave function expansion for a polaron of zero momentum
\begin{align}\label{varwavefunction}
\ket{\Psi}&=\sqrt{Z}\ket{BEC}\ket{\vecp=0}_I+\sum_\veck \alpha_\veck \bed_{-\veck} \be_\mathbf{0}\ket{BEC} \ket{\veck}_I\nonumber\\
&+\sum_{\veck\vecq} \alpha_{\veck\vecq} \bed_{-\veck} \bed_{\veck-\vecq} \be_\mathbf{0}\be_\mathbf{0}\ket{BEC} \ket{\vecq}_I+\ldots
\end{align}
where $\ket{\vecp}_I$ denotes impurity momentum states. The unperturbed impurity-bath state in the first terms becomes supplemented by `particle-hole' fluctuations of the BEC state given by the second, etc. terms. These terms describe the polaronic dressing of the impurity by bath  excitations that leads to the formation of the polaron. A Fourier transformation of the parameter $\alpha_\veck$ to real space reveals the creation of density modulations in the medium, which represent the formation of a dressing cloud that is build around the impurity particle.
 In fact, wave functions of the type in Eq.~\eqref{varwavefunction} were found to yield a rather accurate description of impurities coupled to a fermionic bath of cold atoms via Feshbach resonances \cite{Chevy2006,massignan_repulsive_2011,Punk2009,Cui2010,Schmidt2012b} and also describe  exciton-impurities in two-dimensional semiconductors \cite{Sidler2016}. When truncated at the one-excitation level, such an ansatz leads  to limited agreement with experimental observations for impurities in a Bose gas \cite{Li2014,ssg16}. 

Dressing can also be understood from a perturbative expansion in terms of Feynman diagrams as shown in Fig.~\ref{fig:Feynman}. Here solid lines represent the propagating impurity, and the dashed lines denote  Bosons that are excited from the BEC. The Fr\"ohlich model considers only processes where bosons are excited out of the condensate and then re-enter  the BEC in their next scattering event with the impurity.

However, these  `Fr\"ohlich scattering processes' can neither account for molecular bound-state formation, nor for intricate strong-coupling physics found close to Feshbach resonances. To describe these phenomena, the last term in Eq.~\eqref{appExtFroehlich1} has to be considered. This term allows for scattering of bosons where an excited boson does not directly reenter the BEC but can scatter off the impurity arbitrarily many times. In this process,  illustrated as gray disk in Fig.~\ref{fig:Feynman}, the boson changes its momentum. The infinite repetition of such scattering  processes represents the Lippmann-Schwinger equation in terms of Feynman diagrams and accounts for molecular  bound state  formation.

\section{Rydberg Polaron Hamiltonian}

We now specialize to the case of Rydberg impurities. When a bath atom is excited into a Rydberg state of principal quantum number $n$, it interacts with the surrounding ground-state atoms through a pseudopotential, first proposed by Fermi \cite{fer34}, in which molecular binding occurs through frequent scattering of the nearly free and zero-energy Rydberg electron from the ground-state atom. In this picture, the Born-Oppenheimer potential for a ground-state atom at distance $\vecr$ from the Rydberg impurity, retaining $s$-wave and $p$-wave scattering partial waves, is given as \cite{fer34,gds00,fab86,ckf02,omo77}
\begin{eqnarray}
V_\text{Ryd}(\vecr)&=&\frac{2\pi \hbar^2}{m_e} a_s|\Psi(\mathbf{r})|^2
 +\frac{6\pi \hbar^2 }{m_e} a_p^3|\overrightarrow{\nabla}\Psi(\mathbf{r})|^2,   \label{RydbergInteraction}
\end{eqnarray}
where we kept explicit factors of $\hbar$ and $\Psi(\mathbf{r})$ is the Rydberg electron wave function, $a_s$ and $a_p$ are the momentum-dependent $s$-wave and $p$-wave scattering lengths, and $m_e$ is the electron mass.
When $a_s<0$, $V(\mathbf{r})$ can support molecular states  with
one or more ground-state atoms bound to the impurity \cite{gds00,bbn09,dad15}.
A Rydberg polaron is formed when the Rydberg impurity is dressed by the occupation of a large number of these bound states in addition to finite momentum states. This process gives rise to an absorption spectrum of a distribution of molecular peaks with a Gaussian envelope, which is the key spectral signature of Rydberg polaron formation.

The large energy scale of the Rydberg molecular states involved in the formation of Rydberg polarons allow us to simplify the extended Fr\"ohlich Hamiltonian Eq.~\eqref{appExtFroehlich1}: the typical energy range of Rydberg molecules is $0.1-10$\,MHz for high quantum number $n$, while the typical energy scale for Bose-Bose interactions is 1-10 KHz. Therefore bosons that are bound to a Rydberg impurity probe momentum scales deep in the particle branch of the Bogoliubov dispersion relation, in a regime where the Bogoliubov factors $u_\vecp=1$ and $v_\vecp=0$. Hence we can neglect the Bose-Bose interactions and Eq.~\eqref{appExtFroehlich1} reduces to the simplified, extended Fr\"ohlich model \cite{sln16}
\begin{eqnarray}\label{appHRydberg1}
\hat H&=& \frac{\hat \vecp^2}{2M}
+\sum_\veck \epsilon_\veck \ad_\veck\a_\veck
+\frac{1}{\cal{V}}\sum_{ \veck\vecq}V(\vecq)e^{-i\vecq \hat \vecR}\ad_{\veck+\vecq}\a_{\veck}\nonumber\\
&=& \frac{\hat \vecp^2}{2M}
+\sum_\veck \epsilon_\veck \bed_\veck\be_\veck
+\rho V(\mathbf 0)  \nonumber\\
&+&\underbracket{\frac{1}{\sqrt{\cal{V}}}\sum_{ \vecq}\sqrt{\rho}V(\vecq)e^{-i\vecq \hat \vecR}\left(\bed_{\vecq}+\be_{-\vecq}\right)}_{\text{Fr\"ohlich interaction}}\nonumber\\
&+&\underbracket{\frac{1}{\cal{V}}\sum_{ \veck\vecq}V(\vecq)e^{-i\vecq \hat \vecR}\bed_{\veck+\vecq}\be_{\veck}}_{\text{extended Fr\"ohlich interaction}}.
\end{eqnarray}
where in the second equivalent expression we expanded the boson operators $\ad_\veck$ around their expectation value $\sqrt{N_0}$ which highlights the connection to the Fr\"ohlich model. In order to describe Rydberg impurities the interaction $V(\vecq)$ is given as the Fourier transform of the Rydberg molecular potential $V(\vecq)=\int d^3r V_\text{Ryd}(\vecr)e^{i\vecq\cdot\vecr}$. The molecular potentials and interacting single-particle wave functions ($\ket{\beta_i}$) are calculated as described in \cite{dad15,ssd16}.  We note that the depletion of a BEC by a single electron in a BEC has been studied in \cite{Balewski2013} including only the linear Fr\"ohlich term in Eq.~\eqref{appHRydberg1}. While such an approach can approximately account for the rate of depletion of the condensate it fails to describe the formation of molecules that is essential for the formation of Rydberg polarons.



\section{Linear-response absorption from quench dynamics}


One way to probe polaron structure and dynamics is absorption spectroscopy. Here one utilizes the fact that before being excited to a Rydberg state, an atom is in its ground state $\ket{5s}$, and its interaction with the surrounding Bosons is negligible. The system is described by the Hamiltonian $\hat H_0$ given by the first two terms in Eq.~\eqref{appHRydberg1}. In contrast, when the atom is in its Rydberg state $\ket{ns}$ the potential $V(\vecq)$ is switched on. In experiments \cite{SrPRL2017}, transitions  between both states are driven by a two-photon excitation.

Within  linear response, the corresponding absorption of laser light at frequency $\nu$  is given by Fermi's Golden rule
\begin{align}\label{AbsorptionFGR1}
\mathcal A(\nu)&= 2\pi\sum_{if}w_i |\bra{f}\hat V_\text{L}\ket{i}|^2\times\,\delta(\nu-(E_i-E_f)).
\end{align}
Here the sum extends over all initial and final states of the impurity plus bosonic bath that fulfill $\hat H_0 \ket{i}=E_i\ket{i}$ and $\hat H \ket{f}=E_f\ket{f}$, where $\hat H$ is given by Eq.~\eqref{appHRydberg1}
 so that the final states $\ket{f}$ are eigenstates in the presence of the strong perturbation from Rydberg-boson interactions. In the states $\ket{i}$, all atoms (including the impurity atom) are in the $\ket{5s}$ state, while in final states $\ket{f}$ the impurity is in the atomic $\ket{ns}$ state.
 The laser operator  that drives the two-photon transition between these two atomic states $\ket{5s}$ and $\ket{ns}$, is given by $\hat V_L\sim \ket{ns}\bra{5s}+h.c.$.

Using the Fourier-representation of the delta function in Eq.~\eqref{AbsorptionFGR1}, one can show that the absorption spectrum follows from \cite{ssd16} (for details see \cite{schmidt2017b})
\begin{eqnarray}\label{AwSt}
\mathcal A(\nu)=2\,\text{Re}\,\int_{0}^\infty dt\, e^{i \nu t}\, S(t)
\end{eqnarray}
with the many-body overlap (also called the Loschmidt echo) $S(t)$.

In the general case of finite temperature, the $w_i$ in Eq.~\eqref{AbsorptionFGR1} are the thermodynamic weights of each initial state  given by the diagonal elements of the  density matrix $\hat\rho_\text{ini}=e^{-\beta H_0}/Z_p$, for sample temperature $k_B T=1/\beta$ and partition function $Z_p$.  For finite temperature the Loschmidt echo is then given by
\begin{eqnarray}\label{StFull}
S(t)&=&\text{Tr}[\hat\rho_\text{ini}\, e^{i \hat H_0 t }e^{-i \hat H t}]
\end{eqnarray}
where $\text{Tr}$ denotes the trace over the complete many-body Fock space.  This expression shows that the overlap function $S(t)$  encodes the non-equilibrium time evolution of the system following a quantum quench represented by the introduction of the Rydberg impurity at time $t=0$. From this time evolution then follows the absorption spectrum by Fourier transformation.

The relevant time scales for the dynamics of Rydberg impurities in a BEC is given by the Rydberg molecular energies which exceed both the temperature scale $k_B T$ and the energy scale associated with boson-boson interactions. Thus temperature  and boson interaction effects can be neglected in the calculation of $S(t)$ for Rydberg polarons. In this limit the initial state of the system is given by $\ket{i}=\ket{\text{BEC}}\otimes \ket{\vecp=0}_I \otimes \ket{5s}_I$ and Rydberg polarons are well described by the T=0 limit of Eq.~\eqref{StFull},
\begin{equation}\label{ST0}
S(t)={_I\bra{\vecp=0}}\bra{\Psi_\text{BEC}}e^{i \hat H_0 t}e^{-i \hat H t}\ket{\Psi_\text{BEC}}\ket{\vecp=0}_I,
\end{equation}
where $\ket{\Psi_\text{BEC}}$ represents an ideal BEC of atoms.

The accurate calculation of the absorption response presents a formidable theoretical challenge. In the following we will present three different approaches of different degrees of sophistication. In a simple approximation one may employ mean-field theory (Section \ref{meanfieldsection}) where the impurity-boson interactions are solely described by the mean-field result $\rho V(\mathbf0)$ in Eq.~\eqref{appHRydberg1}. In this approach all quantum effects {and} fluctuations are neglected. Furthermore one may employ a classical stochastic model (Section \ref{Sec.CMC}), which at least can treat fluctuations to a certain degree. Finally, we employ a functional determinant approach, which was  developed in \cite{ssd16} and which we review in Section VI. This approach treats all interaction terms in Eq.~\eqref{appHRydberg1} fully on a  quantum level and allows us to explain the intricate Rydberg polaron formation dynamics observed in our experiments.

\section{Mean Field Approach}\label{meanfieldsection}

The simplest  treatment of the polaron excitation spectrum is to neglect all interaction terms other than $\rho {V}(\mathbf0)$  in Eq.~\eqref{appHRydberg1}, which yields the mean-field approximation. Hence at a given constant density $\rho(\vecr)$ in the trap the absorption spectrum is obtained by Fourier transformation of (cf. Eq.~\eqref{ST0})
\begin{equation}
S(t,\vec{r})=e^{-i \rho(\vecr) V(\mathbf 0) t} = e^{-i \Delta(\vecr)t},
\end{equation}
which yields a delta function Rydberg-absorption response $\mathcal{A}(\nu,\vecr)=\delta(\nu-\Delta(\vecr))$  for a given local, homogeneous density $\rho(\vecr)$ at a detuning from the atomic transition
\begin{eqnarray}
\Delta(\vecr)&=& \rho(\vecr) V(\vecq=\mathbf 0)=\rho(\vecr)\int d^3\vecr' V(\vecr-\vecr') .
\end{eqnarray}
In the experiment, the density $\rho(\mathbf{r})$ varies with position $\vecr$.  Since the excitation laser illuminates the entire atomic cloud it excites Rydberg atoms in regions of varying density. To theoretically model the resulting average over various contributions from the atomic cloud, we perform a local density approximation (LDA), which assumes the density variation is negligible over the range of the Rydberg interaction $V(\vecr)$ given by Eq.\ \ref{RydbergInteraction}.

The spectrum for creation  of a Rydberg impurity is then given by
\begin{equation}
A(\nu)\propto\int d\vecr^3  \rho(\vecr) \mathcal{A}(\nu,\vecr).
\end{equation}
 In the mean-field approximation this yields the response
\begin{eqnarray}\label{eq:meanfieldspectrum}
A(\nu)\propto \int  d^3r\, \rho(\mathbf{r}) \delta(\nu-\Delta(\mathbf{r}))
\end{eqnarray}
for laser detuning $\nu$ from unperturbed atomic resonance. Note that the mean-field treatment is similar to the description of $1S-2S$ spectroscopy of a quantum degenerate hydrogen gas given in \cite{kil00}.

\begin{figure}[b]
\includegraphics[width=3in,trim=0in 0in 0in 0in]{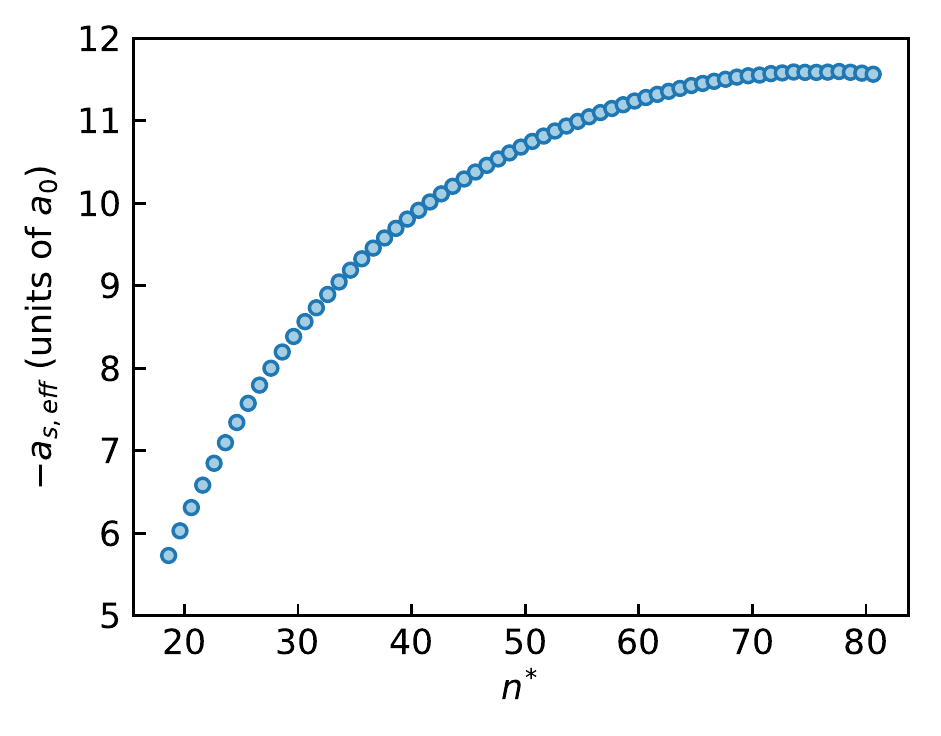}
\caption{Calculated values of $a_{s,\textup{eff}}$ from Eq.~\eqref{eq:meanfieldshift2Aeff} versus $n^*=n-
\delta$, where $\delta=3.371$ is the quantum defect.  }
\label{fig:Aeff}
\end{figure}

An intuitive way to express this shift is in terms of an effective $s$-wave electron-atom scattering length that reflects the average of  the interactions over the Rydberg wave function \footnote{for the remainder of this section we keep factors of $\hbar$},
\begin{eqnarray}\label{eq:meanfieldshift2Aeff}
a_{s,\textup{eff}}=\int  d^3r' \frac{{2\pi m_e}}{h^2}V(\mathbf{r'}),
\end{eqnarray}
yielding
\begin{eqnarray}\label{eq:meanfieldshift3}
h\Delta (\mathbf{r})&=& \frac{h^2 a_{s,\textup{eff}}}{2\pi m_e} \rho(\mathbf{r}).
\end{eqnarray}
$a_{s,\textup{eff}}$ varies with principal quantum number through the variation in the Rydberg electron wavefunction and the classical dependence of the electron momentum on position. Figure~\ref{fig:Aeff} shows calculated values of $a_{s,\textup{eff}}$ for the interaction of Sr($5sns$\,$^3S_1$) Rydberg atoms with background strontium atoms \cite{dad15}.
To obtain this result, Eq.~\eqref{eq:meanfieldshift2Aeff} is only integrated over $|\mathbf{r}'|>0.06 n^2 a_0$ because the approximation breaks down near the Rydberg core, where, among other modifications, the ion-atom polarization potential becomes important.

\begin{figure*}
\includegraphics[width=\textwidth]{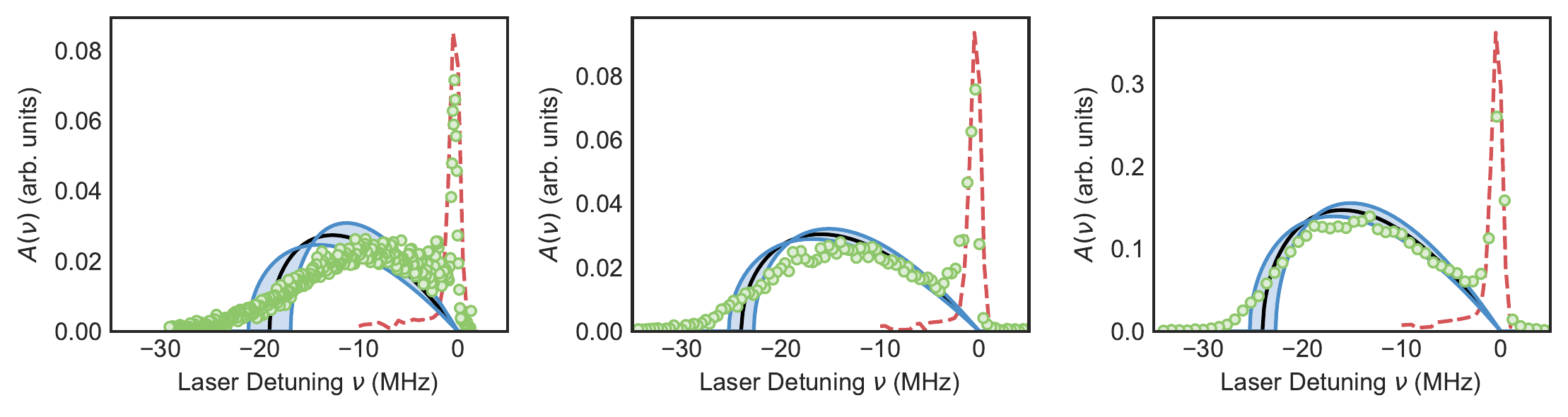}
\caption{Mean-field description of  the excitation spectrum of Sr($5sns$\,$^3S_1$) Rydberg atoms in a strontium  Bose-Einstein condensate (BEC) for (Left) n=49, (Middle) n=60, (Right) n=72. Symbols are the experimental data. Blue bands represent the confidence interval for the mean-field fit of the BEC contribution to the spectrum corresponding to the uncertainties in parameters given in Tab.\ \ref{Tab:FitParams}. The central black line indicates the best fit. {The dashed red line is the mean-field prediction for the
contribution from the low-density region of the atomic cloud formed by thermal atoms. This contribution is calculated using the parameters
given in Tab.\ \ref{Tab:FitParams} including a sample temperature adjusted to reproduce the observed BEC fraction.}
}
\label{Fig:CondensateFraction}
\end{figure*}

\begin{table*}[tbh]
\begin{tabular}{|l|l|l|l|l|l|l|l|l|}
  \hline
  n& $\Delta_{\textup{max}}$ [MHz] & $\eta$ & $N_{BEC}$ & $\bar{\omega}/2\pi$ [Hz] & $\rho_{\textup{max}}~[\textup{cm}^{-3}]$  &  $\mu/k_B$ [nK] & $\Delta_{\textup{max}}/\tilde{\Delta}_{\textup{max}}$ & T [nK]\\
  \hline
   49& $-19\pm2$ & $0.72\pm0.06$ &$(2.8\pm0.3)\times 10^5$& $107 \pm 10$ &$(3.2\pm0.3)\times 10^{14}   $ &$(150\pm 20)$ & $0.95\pm0.1$ & $160\pm 10$   \\
   60& $-24\pm1$ & $0.77\pm0.03$ &$(3.7\pm0.4)\times 10^5$& $112 \pm 10$ &$(3.8\pm0.4)\times 10^{14} $ &$(180\pm 20)$ & $0.93\pm0.05$ & $170\pm 10$ \\
   72& $-24\pm1$ & $0.80\pm0.03$ &$(3.6\pm0.4)\times 10^5$& $117 \pm 10$ &$(3.9\pm 0.4)\times 10^{14}$ &$(190\pm 20)$ & $0.88\pm0.05$ & $170\pm 10$ \\
  \hline
\end{tabular}
\caption{
Parameters for data and fits shown in Fig.~\ref{Fig:CondensateFraction}:  Peak detuning of the mean-field fit ($\Delta_{\textup{max}}$) and BEC fraction ($\eta$) are determined from the spectra. Number of atoms in the condensate ($N_{BEC}$) and mean trap frequency $\bar{\omega}$ are determined from time-of-flight-absorption images and measurements of collective mode frequencies for trapped atoms respectively, and they determine the peak condensate density ($\rho_{\textup{max}}$) and chemical potential ($\mu$). The {second-to-last} column provides the ratio of $\Delta_{\textup{max}}$  to $\tilde{{\Delta}}_{\textup{max}} = \frac{\overline{\omega}}{4\pi}\left(\frac{15 N a_{bb}}{a_{ho}}\right)^{2/5}\frac{m}{m_e}\frac{a_{s,eff}}{a_{bb}}$, the peak shift predicted using information independent of the mean-field fits. {The sample temperature ($T$) is the result of a fit of the observed BEC fraction using a numerical calculation of the number of non-condensed atoms in the trap fixing all the BEC parameters at values given in the table.}
}
\label{Tab:FitParams}
\end{table*}

Figure~\ref{Fig:CondensateFraction} shows the
  mean-field description of experimental results for excitation of Sr($5sns$\,$^3S_1$) Rydberg atoms in a strontium  Bose-Einstein condensate (BEC).

For the prediction of the absorption response the effect of temperature enters by determining the local density of the cloud. In fact, in the local-density-approximation model of the impurity excitation spectrum (Eq.~\eqref{eq:meanfieldspectrum}), the density of atoms in the trap, $\rho(\mathbf{r})=\rho_{BEC}(\mathbf{r})+\rho_{\textup{th}}(\mathbf{r})$, has contributions from thermal, non-condensed atoms and condensed atoms.
The contribution from the thermal gas is restricted to  the sharp peak near zero detuning  and a very small contribution towards the red in each data set in Fig.~\ref{Fig:CondensateFraction}.
Thus the BEC and thermal contributions can be calculated separately in the  mean-field approximation.

The profile of the condensate density is well-approximated for our conditions with a Thomas-Fermi (TF) distribution for a harmonic trap \cite{dgp99}. If we neglect the contribution of thermal atoms to the density,  the condensate contribution to the spectrum is
\begin{equation}\label{eq:thermalg}
  A_{\textup{MF}}(\nu)\propto N_{BEC}\frac{-\nu}{\Delta_{\textup{max}}^2}\sqrt{1-\nu/\Delta_{\textup{max}}},
\end{equation}
for $\Delta_{\textup{max}}\le \nu\le 0$ and zero otherwise \cite{kil00}. Here,
$\Delta_{\textup{max}} =\frac{  h a_{s,\textup{eff}}}{2\pi m_e} \rho_{\textup{max}}$,
and the peak BEC density is
$\rho_{\textup{max}}=\mu_{TF}/g$, where the chemical potential is
$\mu_{TF}=(\hbar \overline{\omega}/2)\left( 15 N_{BEC} a_{bb}/ a_{\textup{ho}} \right)^{2/5}$
and $g=4\pi \hbar^2 a_{bb}/m$
for harmonic oscillator length $a_{\textup{ho}}=(\hbar/m\overline{\omega})^{1/2}$ and mean trap radial frequency $\overline{\omega}=(\omega_1 \omega_2 \omega_3)^{1/3}$. This functional form is used to fit several data sets in  Fig.~\ref{Fig:CondensateFraction} with $\Delta_{\textup{max}}$ and the overall signal amplitude as the only fit parameters. {When frequency is scaled by $\Delta_{\textup{max}}$, the mean-field prediction for a BEC in a harmonic trap is universal, making $\Delta_{\textup{max}}$ an important parameter for describing the spectrum.}
The resulting fit parameters are given in table \ref{Tab:FitParams}. We ascribe the difference between the mean-field fit and the observed spectra at small detuning to the contribution from
the low-density region of the atomic cloud formed by thermal atoms.
From the ratio of the areas of  these two signal components, we extract the condensate fraction. Given the clean spectral separation between the contributions from thermal and condensed atoms, this is a promising technique for measuring very small thermal fractions and thus temperature of very cold Bose gases. {Once the BEC parameters are set, for a consistency check, we then calculate the contribution to the spectrum from thermal atoms using the mean-field approximation and adjusting the sample temperature to match the BEC fraction.} Additional details of the fitting process are described in App. \ref{appendix:BECTempFit}.

For $n=72$, the mean-field approach is quite accurate in describing the overall shape of the response across the entire spectrum. However, at lower principal quantum numbers, the data and fit deviate significantly, especially at larger detunings. Generally, mean-field fails both at
small detuning (i.e. for response from low-density regions of the cloud), where the deviation arises from the formation of few-body molecular states, and at large detuning, where fluctuations in the macroscopic occupation of molecular states become important. Fluctuations correspond to a spread in binding energies of the polaron states excited for a given average density.  All of these effects are neglected in the mean-field description.

\section{Functional determinant approach}\label{FDASection}


The mean-field approach is   insufficient to describe the main features found in the absorption spectrum both at small and large detuning from the atomic transition. For instance its failure at low detuning (low densities) has its origin in the formation of Rydberg molecular states, which is not described in mean-field theory. The  low density regime can be most conveniently studied in detail by working at a low principal quantum number (here $n=38$) where only few atoms are in the Rydberg orbit, and the signal is thus  dominated by the  low-density response. To calculate $S(t)$ efficiently we evaluate it within the FDA.  In \cite{ssd16} the FDA was developed to include (for impurities of infinite mass)  also the finite-temperature corrections that are however irrelevant for our experiment. Thus we can restrict ourselves to the description of the  zero-temperature response given by Eq.~\eqref{ST0}, where the bosonic ground state can be expressed as a state of fixed particle number $\ket{\text{BEC}}=(\hat a_\mathbf{0}^\dagger)^N_0/\sqrt{N_0!}\ket{0}$. For sufficiently large particle number we may equally use its representation as a coherent state $\ket{\text{BEC}}=\exp[{\sqrt{N_0}(\hat a_\mathbf{0}^\dagger-\hat a_\mathbf{0})}]\ket{0}$ where $\ket{0}$ is the boson vacuum. The FDA, as developed in Ref. \cite{ssd16}, did not include the description of impurity recoil which is, however, essential for an accurate prediction of Rydberg molecular spectra. In the following subsection we develop a extension of the FDA approach that overcomes this limitation.

\subsection{Canonical transformation: mobile Rydberg impurity}

The FDA can be applied to time evolutions described by a  Hamiltonian bilinear in creation and destruction operators. This is fulfilled for the Hamiltonian Eq.~\eqref{appHRydberg1} in the case of an infinitely heavy impurity $M=\infty$. However, for a mobile Rydberg impurity, the presence of the non-commuting operators $\hat \vecp$ and $\hat \vecr$ effectively gives  rise to non-bilinear terms.

\begin{figure*}[t]
\includegraphics[width=0.8\textwidth]{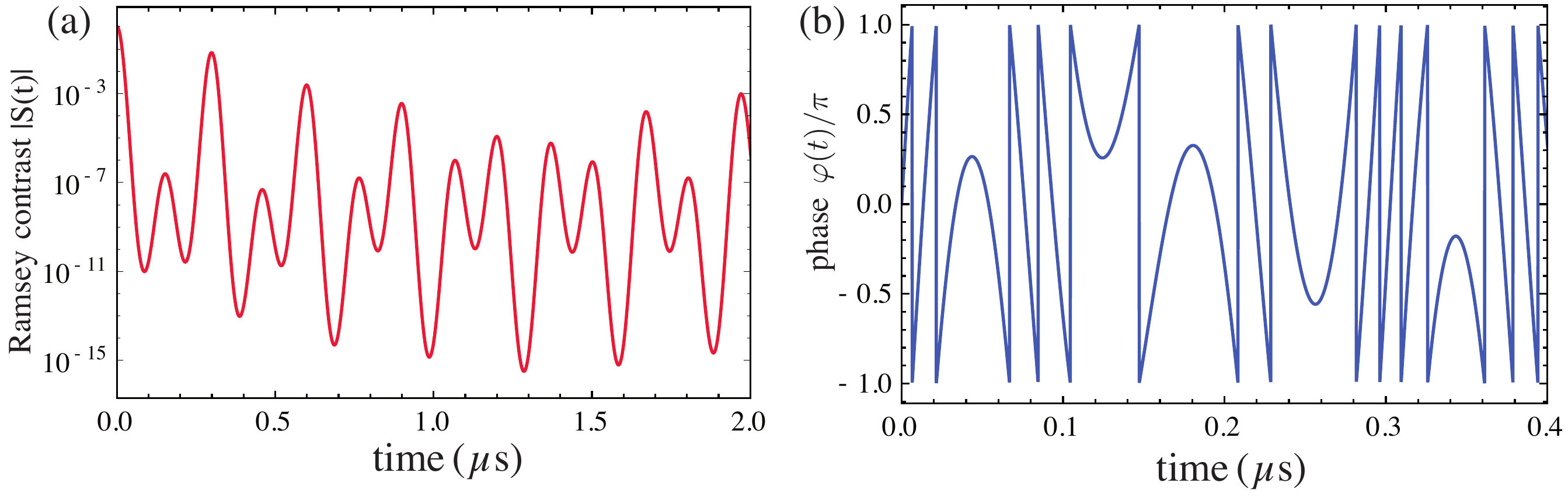}
\caption{Exemplary Ramsey signal $S(t)= |S(t)| e^{i \varphi(t)}$ including contrast $|S|$ and phase $\varphi$ underlying the calculation of the Rydberg polaron absorption spectrum for $n=49$ at peak density in absence of a finite Rydberg lifetime. A  fast decay visible in the contrast accompanies fast oscillations of the full complex system. The combination of both gives rise to the distinct Rydberg polaron features of molecular peaks that are distributed according to a gaussian envelope. The Ramsey signal thus provides  an alternative pathway for observing Rydberg polaron formation dynamics in real-time. The  time scales of this coherent dressing dynamics are ultrafast compared  to the typical time scales of collective low-energy excitations of ultracold quantum gases allowing study of ultrafast dynamics in a new setting. }
\label{fig.Ramsey}
\end{figure*}

To remedy this challenge we combine the FDA with a canonical transformation. Here we focus on the zero-temperature case.
In this case the response is obtained from the time evolution using Eq.~\eqref{appHRydberg1}, where the Hamiltonian includes  bosonic and impurity operators. To deal with the impurity motion we perform a canonical transformation first proposed by Lee, Low, and Pines \cite{Lee1953} which effectively transforms into the system comoving with the impurity. To this end we define the translation operator \cite{Lee1953}
\begin{equation}\label{LLPTrafo}
U=\exp\left\{i \hat \vecR \sum_\veck \veck \ad_\veck \a_\veck \right\}
\end{equation}
which is inserted in the time evolution
\begin{align}\label{LLP2}
S(t)&=\bra{\vecp=0}_I\bra{\Psi_\text{BEC}}U U^{-1}e^{-i \hat H t}U U^{-1}\ket{\Psi_\text{BEC}}\ket{\vecp=0}_I\nonumber\\
&=\bra{\vecp=0}_I\bra{\Psi_\text{BEC}}e^{-i \hat{\mathcal{H}} t} \ket{\Psi_\text{BEC}}\ket{\vecp=0}_I.
\end{align}
In the first line we used that the term $e^{i\hat H_0 t}$ can be dropped as the initial state is a zero energy state. Furthermore, in the second line we made use of the fact that the total boson momentum of the BEC state is zero and we defined the transformed Hamiltonian
\begin{align}\label{FullLLPH}
\hat{\mathcal H}=U^{-1}\hat H U&= \frac{\left(\hat \vecp-\sum_\veck \veck \ad_\veck \a_\veck\right)^2}{2M}
+\sum_\veck \epsilon_\veck \ad_\veck\a_\veck  \nonumber\\
&+\frac{1}{\cal{V}}\sum_{ \veck\vecq}V(\vecq)\ad_{\veck+\vecq}\a_{\veck}
\end{align}
By virtue of the transformation $U$ the impurity coordinate is eliminated in the interaction and only the impurity momentum operator $\hat \vecp$ remains. It thus commutes with the many-body Hamiltonian and is replaced by a c-number, $\hat \vecp\to \vecp$.  Since in our case we are interested in the polaron at zero momentum, $\vecp\to0$. After normal ordering, we arrive at
\begin{eqnarray}\label{LLPTrafo2}
\hat{\mathcal H}&=& \sum_{\veck\veck'} \frac{\veck\veck'}{2M} \ad_{\veck'}\ad_\veck \a_\veck \a_{\veck'}
+\sum_\veck \frac{\veck^2}{2\mu_\text{red}} \ad_\veck\a_\veck   \nonumber\\
&+&\frac{1}{\cal{V}}\sum_{ \veck\vecq}V(\vecq)\ad_{\veck+\vecq}\a_{\veck}
\end{eqnarray}
The above transformation has the effect that the boson dispersion relation $\epsilon_\veck\to \veck^2/2\mu_\text{red}$ now has the reduced mass $\mu_\text{red}=mM/(M+m)$ of boson-impurity partners, and the Hamiltonian includes an induced interaction of bosons described by the first term in Eq.~\eqref{LLPTrafo2}. Due to spherical symmetry and the large energy scale of the Rydberg impurity-Bose gas interaction, we may neglect this term. The reliability of this approximation has been demonstrated  in recent work of some of the authors \cite{ssg16} where a time-dependent variational principle has been applied to the evaluation of the time-evolution of Eq.~\eqref{LLPTrafo2}. In the approximation where the time-dependent wave function  $\ket{\Psi(t)}$ evolved by $\exp(-i\hat{\mathcal{H}}t)$ is taken to be a product of coherent states  of the form $\ket{\Psi(t)}=e^{\sum_\veck (\gamma_\veck(t) \hat a_\veck^\dagger+\text{h.c.})}\ket{0}$, one finds from the equation of motions of the variational parameters $\gamma_\vecq$ that the expectation value $\bra{\Psi(t)}\sum_{\veck\veck'} \frac{\veck\veck'}{2M} \ad_{\veck'}\ad_\veck \a_\veck \a_{\veck'}\ket{\Psi(t)}$ always remains zero and this term hence does not contribute to the dynamics. For our case of Rydberg impurities the FDA solution presented here reproduces the variational result of Ref.~\cite{ssg16} and again shows excellent agreement with experimental data attesting a postiori to the accuracy of the method and the neglect of the first term in Eq.~\eqref{LLPTrafo2}. To which extent this remains valid for other Bose polaron scenarios is an open question and subject to ongoing theoretical studies \cite{GrusdtSchmidt2017}. In summary, we finally arrive at the Hamiltonian
\begin{equation}\label{FinalH}
\hat{ \mathcal H}= \sum_\veck \frac{\veck^2}{2\mu_\text{red}} \ad_\veck\a_\veck
+\frac{1}{\mathcal V}\sum_{ \veck\vecq}V(\vecq)\ad_{\veck+\vecq}\a_{\veck} \ket{ns}\bra{ns}
\end{equation}
whose dynamics we simulate to obtain Rydberg polaron excitation spectra. Note, in Eq.~\eqref{FinalH}, we make explicit that the impurity-boson interaction is only present when the impurity is in its excited atomic Rydberg state  $\ket{n s}$.  In this state the Rydberg electron scatters off ground state atoms in the environment, leading to the strong Rydberg Born-Oppenheimer potential $V(\vecq)$. For simplified notation we will now switch again to the symbol $\mathcal{\hat H} \to \hat H$.

\subsection{Time-domain $S(t)$}

The strength of the FDA is that it allows one to express overlaps of many-body states in terms of single-particle eigenstates \cite{schoenhammer07,abanin_fermi-edge_2005,klich_03,levitov_electron_1996,ssd16}. In our case this implies that we must calculate single particle eigenstates and energies of the free and `interacting' Hamiltonian, i.e. $\hat h_0\ket{n}=\epsilon_n \ket{n}$ and $\hat h\ket{\beta} = \omega_\beta \ket{\beta}$, respectively ($\hat h_0$ and $\hat h$ are the single-particle representatives of the many-body operators $\hat H_0$ and $\hat H$, respectively).
The single-particle eigenstates and energies  $\ket{\beta}$  and $\omega_\beta$ are calculated using exact diagonalization.
Here we solve the radial Schr\"odinger equation for a spherical box of radius $R$ discretized in real-space. Since the BEC is initially in a zero angular state and the Rydberg molecular potential is spherically symmetric we can restrict the analysis to single-particle states with zero-angular momentum. We include the 300 energetically lowest eigenstates, which leads to convergent results.
In terms of these states and energies the overlap $S(t)$ becomes \cite{ssd16}
\begin{equation}\label{ZeroTS}
S(t)=\left(\sum_{\beta} |\braket{\beta}{s}|^2 e^{i(\epsilon_s-\omega_\beta)t}\right)^{N},
\end{equation}
where $\ket{s}$ denotes the lowest single particle eigenstate \cite{statenote}
of $\hat h_0$, and $N$ is the number of atoms in the spherical box of radius $R$ chosen to reproduce a given {local} experimental density of a Bose gas. We choose $R=10^6 a_0$ with $a_0$ the Bohr radius so that we find negligible finite size corrections.
We emphasize again that the temperature enters the calculation only in determining the local density of the atomic cloud. For the dynamics and thus the prediction  of  the absorption spectra at a given {local} density the temperature $T$ is irrelevant. This is due to the fact that the energy scales of the dynamics of the system is determined by the binding energy of Rydberg molecular states, which exceed $k_BT$.



We obtain  absorption spectra by predicting the many-body overlap, or Loschmidt echo, $S(t)$. This calculation corresponds to solving the full time evolution of the system following a quantum quench, where at time $t=0$ the Rydberg impurity is suddenly introduced into the BEC.   The overlap $S(t)$  describes the dephasing dynamics of the many-body system and thus the evolution of the dressing of the Rydberg impurity by bath excitations. It is one of the virtues of cold atomic systems that the  signal $S(t)$ can be directly measured experimentally by Ramsey spectroscopy where via a $\pi/2$ rotation at time $t=0$, the impurity is prepared  in a superposition state of $\ket{5s}$ and $\ket{ns}$. Following a time evolution of duration $t$, a further $\pi$ rotation is performed and  $\sigma_z$ is measured. This gives the Ramsey contrast $|S(t)|$. Changing the phase of the final $\pi$ rotation, the full signal $S(t)= |S(t)| e^{i \varphi(t)}$ can be measured \cite{knap2012,cjl16,schmidt2017b}. In Fig.~\ref{fig.Ramsey} we show the predicted Ramsey signal  that underlies  the calculation of the Rydberg polaron absorption spectrum for $n=49$ at peak density, here shown in absence of a finite Rydberg lifetime.  We observe a  fast decay in the contrast that indicates strong dephasing and thus efficient creation of polaron dressing by particle-hole excitations. This decay is accompanied by fast oscillations of the  complex signal $S(t)$ as visible in the shown evolution of the Ramsey phase $\varphi(t)$ restricted to the branch $(-\pi,\pi)$. The combination of the oscillations at the Rydberg molecular binding energies and the decay of the Ramsey signal $S(t)$ gives rise
to the distinct Rydberg polaron features of molecular peaks that are distributed according to a gaussian envelope, to be discussed below. The Ramsey signal thus provides an alternative pathway to observing Rydberg polaron formation dynamics in real-time. The   time scales of this coherent dressing dynamics are ultrafast compared  to the typical time scales of collective low-energy excitations of ultracold quantum gases. This again highlights that effects arising from Bose-Bose interaction as well as finite temperature will play only a minor role in the prediction of the absorption response as those start to influence dynamics only on much longer time scales.

\begin{figure}[t]
\includegraphics[width=0.97\linewidth]{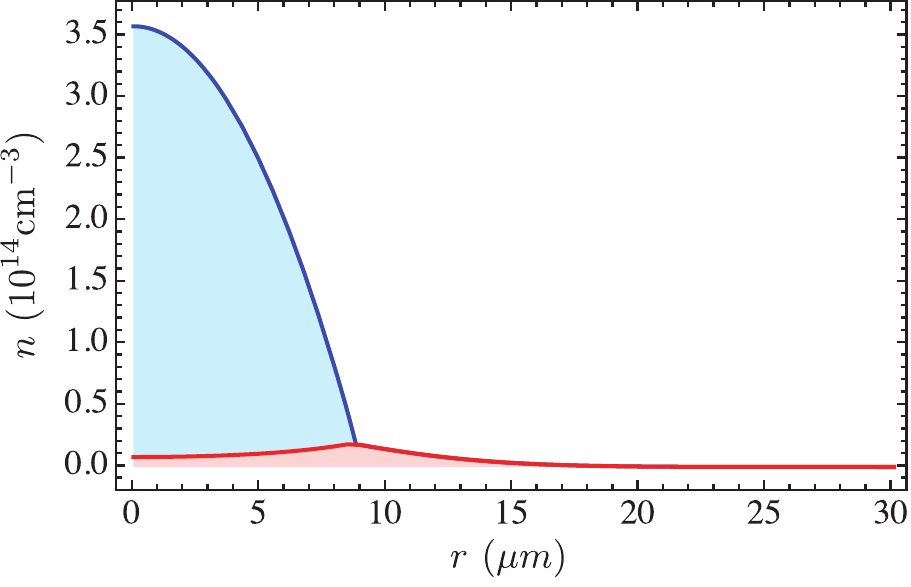}
\caption{Density profile of bosonic atoms in a harmonic trap for parameters $T=180$ nK, $N_\text{BEC}=3.7 \cdot10^5$, $N_\text{tot}=5.2\cdot 10^5$, $\omega_r = 104$ Hz, and $\omega_a=111$ Hz, used for the prediction of the spectrum of the $n=60$ Rydberg excitation shown in Fig.~\ref{fig.Spec60}. In blue is shown the contribution from the condensed atoms, while the red area represents the contribution from thermal atoms. }
\label{fig.HFProfiles}
\end{figure}

\begin{figure}[b]
\includegraphics[width=\linewidth]{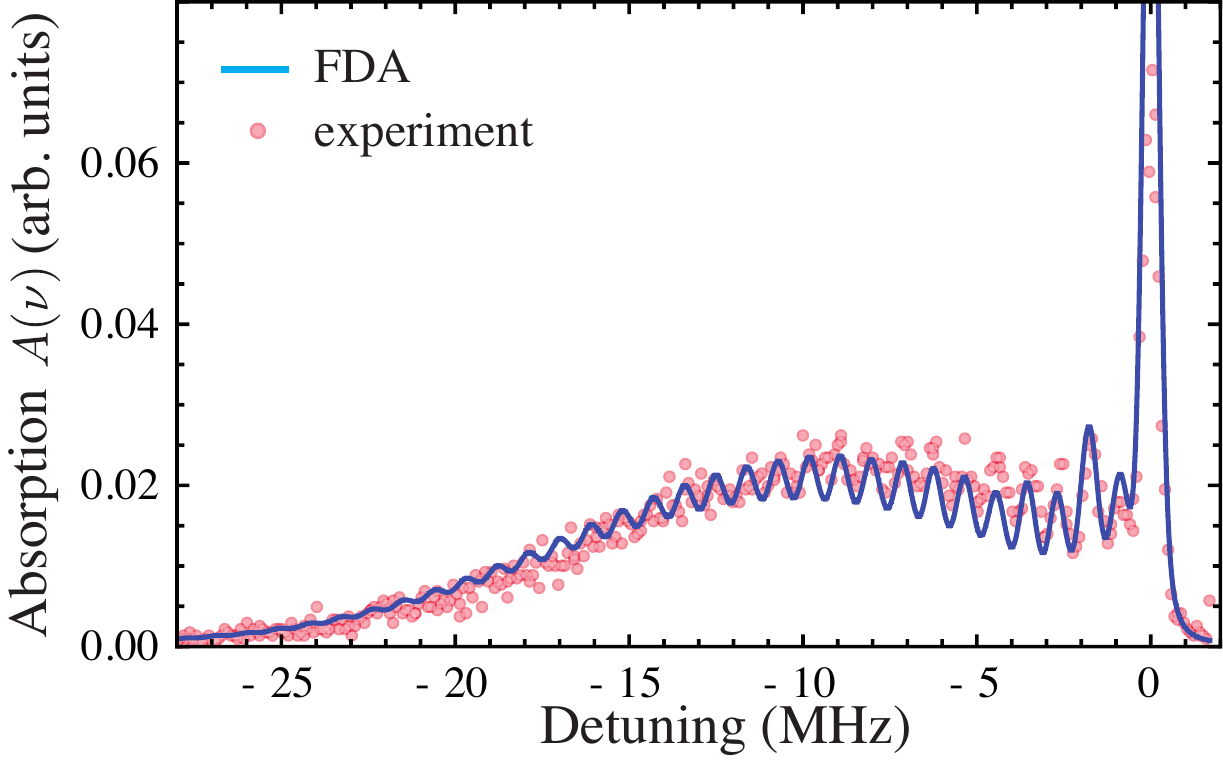}
\caption{(a) LDA absorption spectrum for $n=49$ as calculated by FDA (solid blue) in comparison with experimental data (symbols). }
\label{fig.CMCvsFDT49}
\end{figure}

Using the data of the non-equilibrium quench dynamics, the FDA can accurately capture the formation of Rydberg molecular dimers, trimers, tetramers, etc. This is demonstrated by comparing the FDA spectrum and experimental measurement for low principal quantum number, as shown in \cite{SrPRL2017}.
Indeed, from a many-body wave-function perspective  capturing the trimer, tetrameter and higher-order oligamer state requires the inclusion of the corresponding higher order terms in Eq.~\eqref{varwavefunction}. This attests to the challenge of describing Rydberg polarons which are formed by the dressing with many deeply bound atoms, rather than just a small number of bath excitations. In fact due to the magnitude of the binding energies involved, coherent formation of Rydberg polarons takes place on a $\mu$s timescale, which is short compared to typical ultracold-atom time scales.

\subsection{FDA absorption spectra}


The accurate description of molecular formation underlies the precise analysis of Rydberg-polaron absorption response. As the principal quantum number is  increased to $n=49$, more atoms are situated on average within the Rydberg orbit and many-body effects become relevant. The density-averaged prediction using FDA is shown in Fig.~\ref{fig.CMCvsFDT49}. For all density-averaged spectra, we rely on density profiles such as shown in Fig.~\ref{fig.HFProfiles} for parameters used for the FDA prediction of the $n=60$ Rydberg-absorption spectrum. We include Hartree-Fock corrections \cite{dgp99,smi16} as discussed in Appendix \ref{appendix:BECTempFit}. Furthermore, the absorption spectrum is calculated from the Fourier transformation according to Eq.~\eqref{AwSt}, where in the time-evolution of $S(t)$ we choose a temporal cutoff determined by the experimental laser-pulse duration, and in the Fourier transform we account for the finite laser line width (400 kHz).

\begin{figure}[t]
\includegraphics[width=\linewidth]{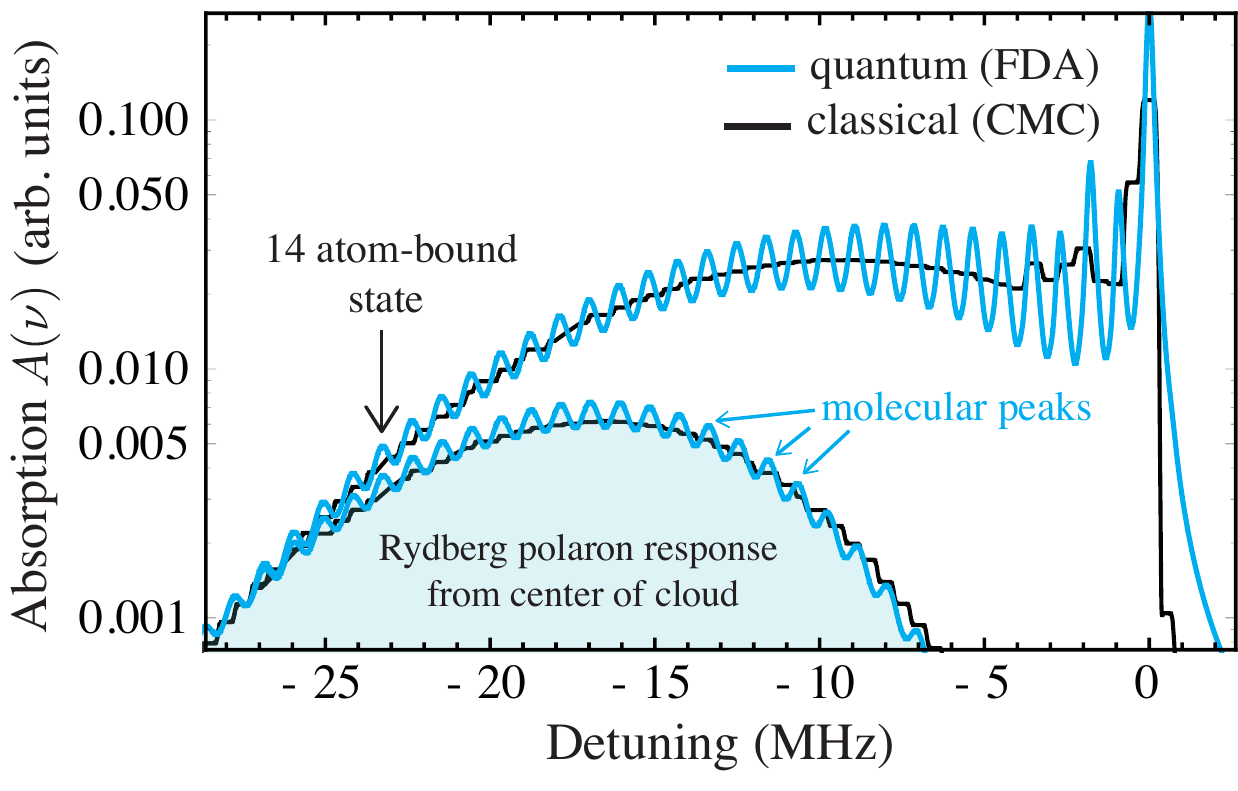}
\caption{LDA absorption spectrum for  $n=49$ as calculated from the classical Monte Carlo sampling model (black) compared to the FDA prediction. The shaded area, which corresponds to the prediction of absorption response from the cloud center at peak density,   reveals that the Gaussian line shape of the response is due to discrete, Gaussian-distributed molecular peaks and not a continuous response as described by a simple classical model.}
\label{fig.CMCvsFDT49detail}
\end{figure}

In Fig.~\ref{fig.CMCvsFDT49detail}, we show again the $n=49$ absorption spectrum obtained theoretically, but with a simulated laser linewidth of 100 kHz (solid blue line). The shaded region  shows the result of an FDA calculation of the spectrum for a central region of the atomic cloud with roughly constant density. We find that the signal from this region shows a series of molecular peaks whose weights follow a Gaussian envelope. This distribution of molecular peaks is one of the key signatures of Rydberg polarons as predicted theoretically in the accompanying work \cite{ssd16}.
The comparison of experiment and theory in Fig. 6 in turn shows that the response is indeed composed of many individual molecular lines. While due to the finite lifetimes of Rydberg molecules and the fact that their binding energies decrease for increasing principal number, these individual molecular peaks cannot  be resolved experimentally for high principal numbers, their Gaussian envelope is a robust signature of the formation of Rydberg polarons.

As our simulation shows, the  broad tail to the red and the characteristic gaussian profile of the signal are both clear signatures of Rydberg polarons. The excellent agreement between many-body theory and experiment  confirms the presence of Rydberg polarons in the Sr experiment.

\begin{figure}[t]
\includegraphics[width=0.97\linewidth]{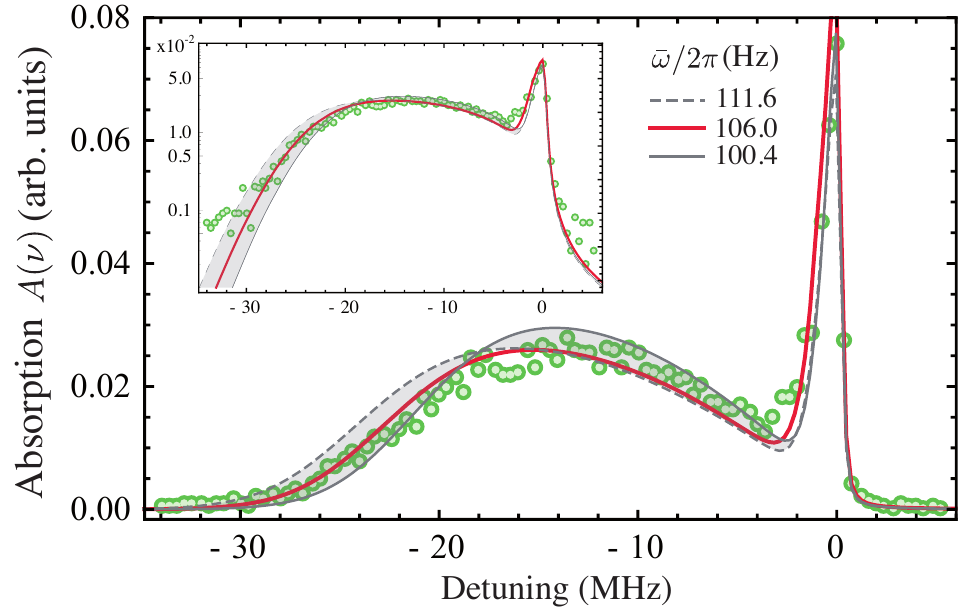}
\caption{Experimentally observed absorption spectrum (symbols) for  $n=60$ in comparison with the theoretical prediction from FDA (lines). The experimental input parameters such as the trap frequencies are varied within the experimental uncertainty, giving rise to the gray band around the solid red line. The specific set of values is given in Table \ref{tableFDT}. The inset shows the signal on a logarithmic scale.
}
\label{fig.Spec60}
\end{figure}

\begin{table*}[t]
\begin{tabular}{|l|l|l|l|l|l|l|l|l|l|}
  \hline
  Data & $\bar{\omega}/2\pi$ [Hz] & ${\omega_i}/2\pi$ [Hz] &  $\eta$ & $N_{BEC}$ &  $N_{tot}$& $\rho_{peak,BEC}$ & $\rho_{peak,tot}$  &  $\mu/k_B$ [nK] &T [nK] \\
  \hline
 Experiment & $112 $ & $(109,109,117)$ & $0.77$ &$3.7 \times 10^5$& $4.8 \times 10^5$&$3.8 \times 10^{14}$ &- & $170$ &-\\
  Fig.~\ref{fig.HFProfiles}(gray,solid)& $111.6 $ & $(109,109,117)$ & $0.77$ &$3.7 \times 10^5$& $4.8 \times 10^5$&$3.72 \times 10^{14}$ &$3.78 \times 10^{14}$ & $175.5$ &$166$\\
  Fig.~\ref{fig.HFProfiles}(red)& $106.0 $ & $(103.6,103.6,111.2)$ & $0.71$ &$3.7 \times 10^5$& $5.2 \times 10^5$&$3.5 \times 10^{14}$ &$3.57 \times 10^{14}$ & $165$ &$180$\\
    Fig.~\ref{fig.HFProfiles}(gray)& $100.4 $ & $(98.1,98.1,105.3)$ & $0.77$ &$3.7 \times 10^5$& $4.8 \times 10^5$&$3.3 \times 10^{14}$ &$3.58 \times 10^{14}$ & $155$ &$150$\\
  \hline
\end{tabular}
\caption{
Parameters used for the prediction of the theoretical spectrum shown in  Fig.~\ref{fig.HFProfiles}(b) for the Rydberg $n=60$ excitation. Inferred quantities are the condensate fraction $\eta=N_\text{BEC}/N_\text{tot}$, peak BEC density $\rho_\text{peak,BEC}$, and chemical potential $\mu$. Densities are given in $\text{cm}^{-3}$.}
\label{tableFDT}
\end{table*}

The Gaussian signature of the Rydberg-polaron spectrum becomes  more pronounced when exciting $n=60$ states (Fig.~\ref{fig.Spec60}).  In the accompanying work \cite{ssd16} we observe this Gaussian response experimentally for Rydberg polarons of large principal number $n=60$ and $n=72$.  While due to the finite lifetimes of Rydberg molecules and the fact that their binding energies decrease for increasing principal number, these individual molecular peaks cannot  be resolved experimentally for high principal numbers, their Gaussian envelope is a robust signature of the formation of Rydberg polarons. The comparison of experiment and theory in Fig. 6 in turn shows that the response is indeed composed of many individual molecular lines.

The FDA calculation of the absorption spectra takes as input the experimentally determined number of atoms in the condensate $N_\text{BEC}$ and the trap frequencies $\omega_i$. The temperature $T$ and the total number of atoms $N_\text{tot}$ are taken as fit parameters, and results are consistent with their determination using the mean-field model and absorption imaging described in Section~\ref{meanfieldsection}. For the $n=60$ spectrum, we also show
a range of FDA predictions resulting from varying the trap frequencies within experimental uncertainty, which illustrates the impact of these uncertainties.
Table \ref{tableFDT} lists the  parameters used in the theoretical simulation. Note that the requirement to explain the spectrum over the whole frequency regime places tight constraints on the fit parameters $T$, and $N_\text{tot}$.
(For Rydberg polaron response at $n=72$, refer to Ref. \cite{SrPRL2017}).

As discussed in Ref.~\cite{ssd16}, the emergence of the Gaussian response can be understood from a direct, analytical calculation of $S(t)$ in terms of single particle eigenstates and energies. Indeed expanding the sum in Eq.~\eqref{ZeroTS} explicitly in  a multinominal form leads, after Fourier transform, to the expression
\begin{eqnarray}\label{SpecAna}
\mathcal{A}(\nu)&=&N! \sum_{\Sigma n_i = N}\frac{|\braket{\alpha_1}{s}|^{2 n_1}\cdot \ldots \cdot|\braket{\alpha_M}{s}|^{2 n_M}\cdot \ldots}{n_1!\ldots n_M! \ldots} \nonumber \\
&&\times \delta(-\nu+n_1 \omega_1+\ldots +n_M \omega_M+\ldots).
\end{eqnarray}
This expression captures not only the bound states but also continuum states. Expressing the spectrum in this explicit form  again highlights the fact that the actual spectrum is indeed built by $\delta$-peaks corresponding to the many configurations in which bound molecules can be formed and thus dress the impurity. It also explains the emergence of the gaussian lineshape as a limit of the multinominal distribution of delta-function peaks for large particle number $N$. We note that the fact that the spectrum is  built by molecular excitation peaks (of up to hundreds of atoms bound to a single impurity) is  missed by the classical description of the Rydberg polarons discussed in Section~\ref{Sec.CMC}.

We note that while Eq.~\eqref{SpecAna} is appealing as it qualitatively explains  the observed spectral features, it is less useful for quantitative calculations due to the exponential growth of the number of terms in the sum (reflecting the exponential growth   of Hilbert space for increasing particle number). In contrast, the calculation of the time-evolution in Eq.~\eqref{ZeroTS} is numerically efficient (when limited to finite evolution times) and only requires a final Fourier transformation.

\section{From  quantum to classical description of Rydberg absorption response} \label{Sec.CMC}

\begin{figure}[b]
\includegraphics[width=\linewidth]{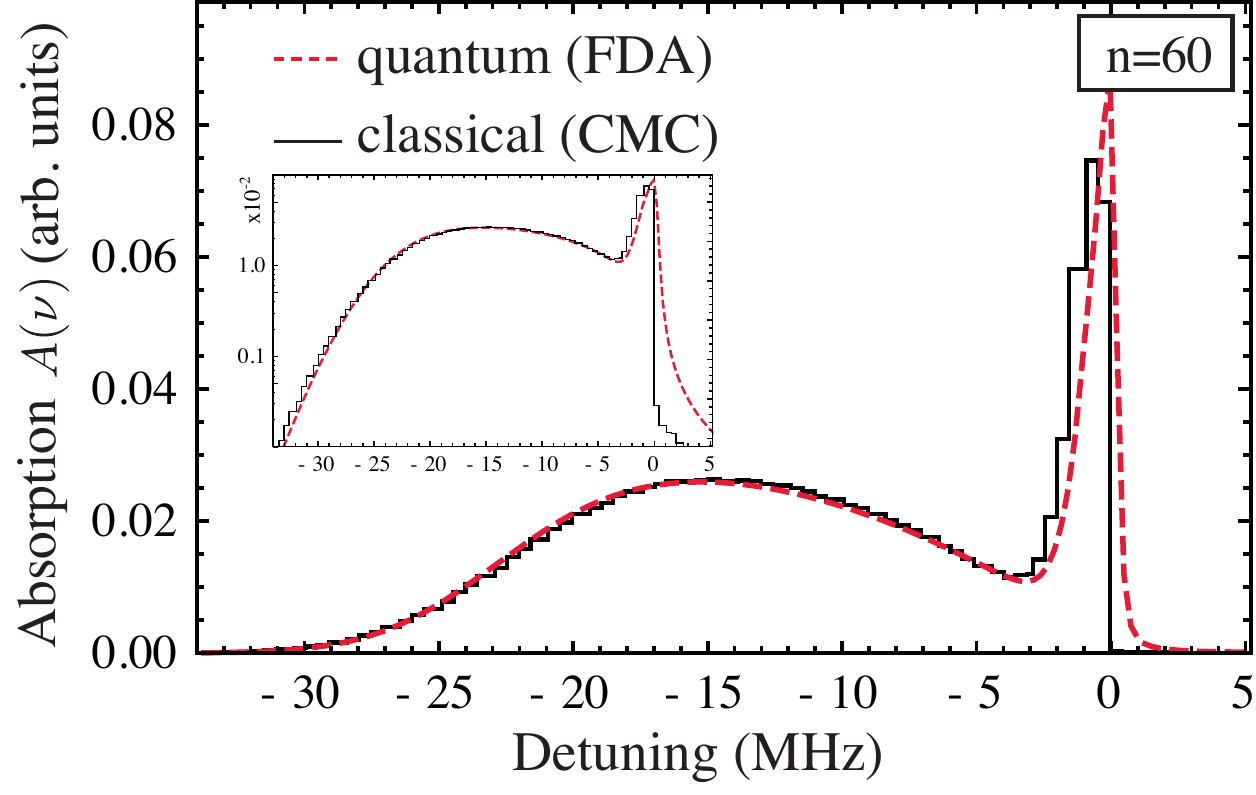}
\caption{LDA absorption spectrum for $n=60$ as calculated by the FDA (dashed red) and classical statistical Monte Carlo sampling method (black). The inset shows the signal on a logarithmic scale.}
\label{fig.CMCvsFDT1}
\end{figure}

For a sufficiently large average number of atoms within the Rydberg electron orbit,
the overall line shape of Rydberg spectra
 can be described with   classical statistical    arguments. In Fig.~\ref{fig.CMCvsFDT1}, we show  the experimental spectrum obtained for $n=60$ in comparison with the prediction from the FDA (dashed red line) and a classical statistical approach  (solid black). In the latter approach, using  a classical Monte Carlo (CMC) algorithm \cite{sln16}, we randomly distribute atoms in three-dimensional space around the Rydberg ion such that the correct density profile is obtained.
Due to statistical fluctuations in the random sampling of coordinates (sampled from a uniform distribution), the local density within the Rydberg-electron radius  fluctuates. For each of the random configurations of atoms $\mathcal C=(\vecr_1, \vecr_2,\ldots)$ we calculate the {classical} energy of the configuration  $E_{\mathcal C}=\sum_i V(\vecr_i)$, where $\vecr_i$ are the atom coordinates.  The resulting energies $E_{\mathcal C}$ are collected in an energy histogram and shown as the black solid line in Fig.~\ref{fig.CMCvsFDT1}.

The   validity of the classical description  reflects the fact that the macroscopic occupation of   the molecular bound and scattering states     probes the Rydberg molecular potential uniformly over its entire range. This process is also well described by a repeated  sampling of different classical atom configurations drawn from a homogeneous density  distribution.

The quantum-classical correspondence also becomes evident when considering the specific approximation that underlies the classical statistical approach. The Rydberg absorption response is given as a Fourier transform of the full quantum quench dynamics as given by Eq.~\eqref{ST0}. The classical statistical model arises as an approximation of the time evolution of $S(t)$ where the kinetic energy of both the impurity and the Bose gas  is completely neglected. Hence both bosons and the impurity are treated as effectively infinitely heavy objects; their motion is `frozen' in space. The disregard of the kinetic terms has the consequence that the Hamiltonian now commutes with the position operators $\hat \vecr_j$ of impurity and bath atoms (non-commuting $\hat \vecp_j$ operators are now absent), and hence dynamics becomes completely classical. In this approximation the Hamiltonian becomes (without loss of generality, we assume that the impurity is at the center of the coordinate system)
\begin{equation}
\hat H= \int d^3r  V(\vecr)\ad(\vecr)\a(\vecr)
\end{equation}
Furthermore, $\hat H_0=0$, and hence $\hat \rho_\text{ini} = 1/Z$ in Eq.~\eqref{StFull}. Since all boson coordinates $\hat \vecr_j$ now commute with the Hamiltonian, the many-body eigenstates are given by $\ket{\phi^{(j)}}=\ket{\vecr^{(j)}_1,\vecr^{(j)}_2,\ldots, \vecr^{(j)}_N}$. The trace in Eq.~\eqref{StFull} reduces to the sum over the set of all basis states $\{\ket{\phi^{(j)}}\}$. The $\ket{\phi^{(j)}}$ are eigenstates of $\hat H$ with eigenvalues $\hat H \ket{\phi^{(j)}}= \sum_{\{\vecr_i^{(j)}\}} V(r_i^{(j)}) \ket{\phi^{(j)}}$ and $S(t)$ becomes
\begin{equation}\label{class1}
 S_\text{cl}(t)=\mathcal N \sum_{\{\vecr_i^{(j)}\}}e^{i\left(\nu - \sum_{i} V(r_i^{(j)})\right)t},
\end{equation}
where the sum extends over {all} possible atomic configurations in real-space, and $\mathcal N$ is a normalization factor (representing the partition function $Z$ in the density operator $\rho_\text{ini}$).
Finally, performing the Fourier transform of $S(t)$ one arrives at
\begin{equation}\label{class2}
\mathcal A(\nu)=\mathcal N \sum_{\{\vecr_i^{(j)}\}}\delta\left(\nu - \sum_{i} V(r_i^{(j)})\right).
\end{equation}
 A finite lifetime of the Rydberg or laser excitation, as well as a finite linewidth of the laser leads to a broadening of the delta function in Eq.~\eqref{class2}. The sum in Eq.~\eqref{class2} is  exactly the object  sampled in the classical Monte Carlo (CMC) approach {derived from first principles}.

Moreover we note that for impurities interacting with an ideal Bose gas with contact interactions, the gaussian spectral signature of polarons \cite{ssd16}  would remain, while in {the} classical   model, a sharp excitation at the atomic transition frequency  {is expected}. Indications of  a Gaussian response have been seen in a recent experiment performed independently at Aarhus \cite{jws16} and   JILA \cite{hvk16}, in agreement with theory \cite{ssg16}.  The classical statistical approach   fails {to describe the} Bose polarons close to a Feshbach resonance  \cite{hvk16,jws16}. Here the impurity indeed interacts with the bath via a contact interaction, and, due to a diverging scattering length, the single bound state present in the problem is highly delocalized and extends in size far beyond the range of the potential. This effect cannot be captured by a classical approach.

While Rydberg  absorption spectra find an {effective}, yet approximate description in terms of classical statistics, such an approach does not reveal the  {quantum mechanical origin} of Rydberg polarons. The classical sampling model treats both the Rydberg and the ground state atoms as infinitely heavy objects.
Zero-point motion  and the discrete nature of the bound energy levels are absent in this treatment, so that it cannot describe the  formation of Rydberg molecular states. This short-coming of the classical approach is evident in Fig.~\ref{fig.CMCvsFDT49detail}, where we show the absorption spectrum for $n=49$.  The FDA (solid blue)
fully describes the quantum mechanical formation of molecules (dimers, trimers, tetrameters, excited molecular states, etc.), which gives rise to  discrete  molecular peaks visible in the spectrum.  In contrast, and as evident from Fig.~\ref{fig.CMCvsFDT49detail}, the existence of molecular states, which are the quantum mechanical building block of Rydberg polarons, is not described by the classical approach (solid black).
 The fact that the classical model can only describe the envelope of the Rydberg polaron response but not the underlying distribution of molecular peaks, is further emphasized by the comparison of FDA and CMC simulation from the center of the atomic cloud shown as shaded region in Fig.\ 7.

\section{Conclusion}

In this work, we have detailed the descriptions of polarons as encountered in condensed matter and in ultracold atomic systems. We time-evolve an extended Fr\"{o}hlich Hamiltonian as relevant for Rydberg excitations, Eq.~\eqref{appHRydberg1}, unitarily to obtain the overlap function, Eq.~\eqref{StFull}, whose Fourier transform {leads to}   the spectral function for Rydberg impurity excitation in a Bose gas, Eq.~\eqref{AwSt}.  We show how different approximations to the  many-body quantum description, such as  mean field and classical Monte Carlo treatments can be derived. Here we extend the bosonic functional determinant approach to the Rydberg polaron problem to account for recoil of the impurity. We show that the FDA approach can correctly  and accurately account for the few-body molecular bound-state formation, as well as the macroscopic occupation of many-body bound and continuum states. The various treatments are compared with experimental data for Rydberg excitation in a $^{84}$Sr BEC \cite{SrPRL2017} with the FDA results reproducing the observed data over a wide range of Rydberg line spectral intensities.

 Rydberg polarons are exemplified on the one hand by their large energy scales, ~1-10 MHz, that allow for coherent polaronic dressing on correspondingly short time scales, and on the other hand large spatial extent, $~1\,\mu$m, leading to large-scale variations of the density in the many-body medium. Another key feature of Rydberg polarons is the coherent dressing of the quantum impurity not by collective low-energy excitations, but by a large number of molecular bound states. This is a new dressing mechanism,  {distinguishing them} from polarons    {encountered} so far in the solid state physics context.

Our FDA time domain analysis suggests a way to investigate the dynamical formation of the Rydberg polarons. A natural extension  {will be} to explore the feasibility of observing Pauli blocking in a degenerate Fermi atomic gas with  non-trivial spatial correlations. Note that one can study polarons with nonzero momentum by  keeping all terms  in Eq. \eqref{FullLLPH} and allowing $\vecp$ to be finite. The effective polaron mass can then be analyzed as previously demonstrated \cite{Shashi2014}.




\textbf{Acknowledgements}:
Research supported by the AFOSR (FA9550-14-1-0007), the NSF (1301773, 1600059, and 1205946), the Robert A, Welch Foundation (C-0734 and C-1844), the FWF(Austria) (P23359-N16, and FWF-SFB049 NextLite).  The Vienna scientific cluster was used for the calculations.  H.~R.~S. was supported by a grant to ITAMP from the NSF.
R.~S. and H.~R.~S. were supported by the NSF through a grant for the Institute for Theoretical Atomic, Molecular, and Optical Physics at Harvard University and the Smithsonian Astrophysical Observatory.  R.~S. acknowledges support from the ETH Pauli Center for Theoretical Studies. E.~D. acknowledges support from Harvard-MIT CUA, NSF Grant (DMR-1308435), AFOSR Quantum Simulation MURI, the ARO-MURI on Atomtronics, and support from  Dr.~Max R\"ossler, the Walter Haefner Foundation and the ETH Foundation. T.~C.~K acknowledges support from Yale University during writing of this manuscript.

\appendix
\section{Determination of BEC parameters and Mean-Field description of the Impurity Excitation Spectrum in a Strontium BEC} \label{appendix:BECTempFit}

The mean-field approximation neglects fluctuations in the density around the Rydberg impurity, which correspond to a spread in binding energies of the polaron states excited for a given average density. This explains the discrepancy between data and fit at large detuning. But as shown in \cite{SrPRL2017}, the broadening resulting from these fluctuations is proportional to $\sqrt{\rho}$ and vanishes as detuning approaches zero. Thus, we assume that the difference between data and the mean-field fit in Fig.\ \ref{Fig:CondensateFraction} for small detuning ($\nu/\Delta_{\textup{max}}<0.5$) arises from non-condensed atoms.

We adjust the area of the mean-field BEC contribution so the sum of the non-condensed and mean-field-BEC signal matches the total experimental spectral area. Because the deviations between the mean-field fit and the BEC spectrum are significant, the fit of the peak shift $\Delta_{\textup{max}}$ is less rigorous. We adjust it to qualitatively match the data and so that the mean-field fit and the data have approximately equal area for ($\nu/\Delta_{\textup{max}}>0.5$). This uncertainty in the fitting procedure decreases with increasing principal quantum number as the experimental data converges towards the mean-field form. In the mean-field description we neglect the laser linewidth of 400\,kHz in this analysis.

The fit parameters are given in Table \ref{Tab:FitParams}. The condensate fraction $\eta$ is directly determined from the spectrum. The uncertainties correspond to values calculated for the extremes of the confidence intervals shown as bands in Fig.~\ref{Fig:CondensateFraction}, except for uncertainties in $\omega_i$ and total atom number, which reflect uncertainties from the independent procedures for measuring these quantities. The fit value of the peak shift,  $\Delta_{\textup{max}}$, can be compared to the  predicted peak shift, $\tilde{\Delta}_\textup{max}$, calculated from independent, \textit{a priori} information:   the number of atoms in the condensate $N_{BEC}$ determined from time-of-flight-absorption images,  the trap oscillation frequencies $\omega_i$ determined from measurements of collective mode frequencies for trapped atoms, and the value of $a_{s,\textup{eff}}$ found theoretically from $V_{Ryd}(\mathbf{r})$. The values of $\Delta_{\textup{max}}$ are all about 10\% below $\tilde{\Delta}_\textup{max}$. This may point to something systematic in our analysis procedure, but it is also a reasonable agreement given our uncertainty in determination of $\omega_i$ and total atom number. The calculation of $a_{s,\textup{eff}}$ from $V_{Ryd}(\mathbf{r})$ (Eq.~\eqref{eq:meanfieldshift2Aeff}) could also account for some of this discrepancy through approximations made to describe the Rydberg-atom potential at short range. Residual deviations in the description of the short-range details of the Rydberg molecular potential lead, for instance, also to the minor discrepancies between the FDA and CMC simulation of the Rydberg response from the center of the atomic cloud, shown in Fig.~\ref{fig.CMCvsFDT49detail}.

For a non-interacting gas in a harmonic trap, the temperature is easily found from $\eta$, and $\bar{\omega}$ \cite{psm02}, and the standard expressions imply a temperature of approximately 240\,nK for the data presented in Fig.~\ref{Fig:CondensateFraction}. However, the situation is much more complicated than this for an interacting gas. There are corrections to the condensate fraction that are independent of the trapping potential, and these are often discussed in terms of the shift of the critical temperature for condensation \cite{dgp99,lhy57,sct11}, but these are all small in our case, reflecting the smallness of relevant expansion parameters: $\rho_{\textup{max}}a_{bb}^3=10^{-4}, a_{bb}/\lambda_{\textup{th}}=10^{-2}$ at 100\,nK, and $a_{\textup{ho}}/R_{TF}=10^{-1}$, where $\lambda_{\textup{th}}$ is the thermal de Broglie wavelength and $a_{\textup{ho}}$ is the trap harmonic oscillator length. For a gas trapped in an inhomogeneous external potential $V_{ext}(\textbf{r})$,
the Hartree-Fock approximation
(as described in detail in Refs. \cite{pit03,psm02})
yields a mean-field interaction between thermal and BEC atoms that creates an effective potential for thermal atoms, $V_{\textup{eff}}(\textbf{r})= V_{\text{ext}}(\textbf{r})+2g[\rho_{BEC}(\textbf{r})+\rho_{\textup{th}}(\textbf{r})]$ that is of a ``Mexican hat" shape rather than parabolic
\cite{psm02,pit03,tsc11}.
This effect is particularly important when the sample temperature is close to or lower than the chemical potential, which is the case here. It increases the volume available to non-condensed atoms, and implies a lower sample temperature for a given value of $\eta$ compared to the non-interacting case.  In other words, the number of thermal atoms can significantly exceed the critical number for condensation predicted for an ideal gas \cite{tsc11,smi16}. 240\,nK is thus an  upper limit of the sample temperature. Sample temperatures extracted from bimodal fits to the time-of-flight absorption images are 100-190\,nK.

Both the mean-field and the FDA fits use a local density approximation for the spectrum. Here the temperature and interaction between bosons enter by determining the precise form of the overall density distribution of the atomic gas. As discussed above, the density profile has contributions from both the condensed and thermal, non-condensed atoms
\begin{equation}
\rho(\vecr)= \rho_\text{BEC}(\vecr) +\rho_\text{th}(\vecr).
\end{equation}
The BEC density $\rho_\text{BEC}$ and resulting chemical potential are assumed to be given by the Thomas-Fermi expressions  \cite{pit03}
\begin{equation}
\rho_{BEC}(\vecr)= \frac{m}{4\pi \hbar^2 a_{bb}}[\mu_{TF}-V_{ext}(\vecr)],
\end{equation}
with $\mu_\text{TF}=\hbar \overline{\omega}/2(15 N_\text{BEC} a_{bb}/a_{ho})^{2/5}$.
$N_\text{BEC}$ is the total number of condensed atoms,  $a_{bb}=123 a_0$ for $^{84}$Sr, and $a_{ho}=(\hbar/m\overline{\omega})^{1/2}$ is the harmonic oscillator length. (In this section we make explicit factors of $\hbar$ and $k_B$.)

The density distribution for atoms in the thermal gas can be calculated from known trap and BEC parameters using a Bose-distribution
\begin{equation}\label{thermalrho}
\rho_\text{th}(\vecr)= \frac{1}{\lambda_T^3}g_{3/2}\left\{e^{-\frac{1}{k_BT}\left[V_\textup{eff}(\vecr)-\mu_{TF}\right]}\right\},
\end{equation}
where  $\lambda_T=\sqrt{\frac{2\pi \hbar^2}{m k_BT}}$ is the thermal wavelength and $g_{3/2}(z)$ the polylogarithm $\text{PolyLog}[\frac{3}{2};z]$ \cite{pit03}, and the chemical potential is set to $\mu_{TF}$.
The parameters
 $T$ and $\mu_{TF}$ are varied self-consistently to fit that spectrum, which can be seen as matching the experimentally observed
total particle number $N_\text{tot}=N_\text{BEC}+N_\text{th}$ and BEC fraction $\eta=N_\text{BEC}/N_\text{tot}$.

The mean-field calculations use the full geometry of the external trapping potential as determined by the optical-dipole-trap lasers. Integrals involving $V_\text{ext}(\vecr)$ are evaluated numerically.
This procedure yields temperatures of $155-170\,nK$, in good agreement with other determinations.
The modification of the potential seen by thermal atoms due to Hartree-Fock corrections has an important consequence for our analysis. The peak  shift in the Rydberg excitation spectrum should in principle reflect the peak condensate density plus the density of thermal atoms at the center of the trapping potential. The contribution from thermal atoms would be a significant correction if the mean-field repulsion of thermal atoms from the center of the trap were ignored. But because the sample temperature is close to the chemical potential, the density of thermal atoms is suppressed at trap center and we neglect it in discussions of the peak mean-field shift in the spectrum.

For the FDA simulation, the dipole trap potential is approximated by $V_\text{ext}(\vecr)=m(\omega_r^2 r^2+\omega_z^2 z^2)/2$.
In Fig.~\ref{fig.HFProfiles}(a) we show the density profile along the radial direction for parameters as used for the FDA prediction of the $n=60$ Rydberg absorption spectrum including Hartree-Fock corrections. We emphasize again that the temperature and interaction between bosons are relevant only for the determination of the density profile of atoms. In the simulation of the absorption spectrum, which is obtained within a local density approximation, this density distribution of atoms enters as input. In this simulation, then locally performed for constant density, both temperature effects and boson-interaction effects can be neglected due to the short time scales associated with Rydberg polaron dressing dynamics (that are given by the Rydberg molecular binding energies).



%


\begin{thebibliography}{64}
\bibitem [{sta()}]{statenote}%
  \BibitemOpen
  \href@noop {} {}\bibinfo {note} {Not to be confused with the notation for
  atomic Rydberg states $\ket{ns}$.}\BibitemShut {Stop}%
\end{thebibliography}

%

\end{document}